# FUZZY AIDED APPLICATION LAYER SEMANTIC INTRUSION DETECTION SYSTEM - FASIDS

S.Sangeetha[1] & Dr.V.Vaidehi[2]

[1]Dept. of Computer Science and Engineering, Angel College of Engineering, Tirupur.

[2]Dept. of Electronics Engineering, Madras Institute of Technology, Chennai.

Email: visual.sangi@gmail.com

**ABSTRACT**

The objective of this is to develop a Fuzzy aided Application layer Semantic Intrusion Detection System (FASIDS) which works in the application layer of the network stack. FASIDS consist of semantic IDS and Fuzzy based IDS. Rule based IDS looks for the specific pattern which is defined as malicious. A non-intrusive regular pattern can be malicious if it occurs several times with a short time interval. For detecting such malicious activities, FASIDS is proposed in this paper. At application layer, HTTP traffic's header and payload are analyzed for possible intrusion. In the proposed misuse detection module, the semantic intrusion detection system works on the basis of rules that define various application layer misuses that are found in the network. An attack identified by the IDS is based on a corresponding rule in the rule-base. An event that doesn't make a 'hit' on the rule-base is given to a Fuzzy Intrusion Detection System (FIDS) for further analysis.

## 1. INTRODUCTION

Intrusion detection is the process of monitoring computers or networks for unauthorized entrance, activity, or file modification. Most of the commercially available Intrusion Detection Systems (IDS) work in the network layer of the protocol stack. This paves way for attackers to intrude at various other layers, especially in the application layer namely HTTP. Application layer based IDS blocks the application layer based attacks that network layer intrusion detection system can't block. Firewalls works in network layer and prevents the attacks entering the network through unauthorized port. Some complex threats can enter through authorized port (HTTP 80) and they go undetected. These threats can be detected by Application layer IDS. Misuse detection uses rule based detection that follow a signature-match approach where attacks are identified by matching each input text or pattern against predefined signatures that model malicious activity (Krugel and Toth 2003). The pattern-matching process is time-consuming. Moreover, attackers are continuously creating new types of attacks. These attacks can be detected by the IDS, if it knows about the attack in the form of signatures. Attack signatures can be specified either in a single-line or by using complex script languages and are used in rule base to detect attacks. Because of the continuously changing nature of attacks, signatures and rules should be updated periodically on IDS.

Rule-based Intrusion Detection System (RIDS) looks for specific pattern that are defined as malicious. A non-intrusive regular pattern can be malicious if it occurs several times with a short time interval. The non-intrusive patterns are checked by the fuzzy component of the proposed architecture for a possible attack. The detection rate increases by checking the non-intrusive patterns using the fuzzy component. This thesis proposes a Fuzzy aided Semantic Intrusion Detection System (FASIDS) which aims at designing the above solution for one of the application layer protocols namely HTTP (Hyper-Text Transfer Protocol). Such an HTTP based semantic IDS detects both the header based attacks and payload (which typically consists of HTML and script) based malicious content.

## 2. ARCHITECTURE OF THE FASIDS

The architecture of the Intrusion Detection System is shown in Figure 1. The block diagram shows the order in which the different modules process the incoming HTTP traffic. An HTTP sniffer collects the application-layer HTTP traffic from the network. It is a typical proxy server which listens through the network interface by getting requests from the browser and forwarding the same to the service provider. The response obtained from the web server is also captured by the sniffer before being forwarded to the browser. Captured data is then separated into header and payload parts by session dispatcher.

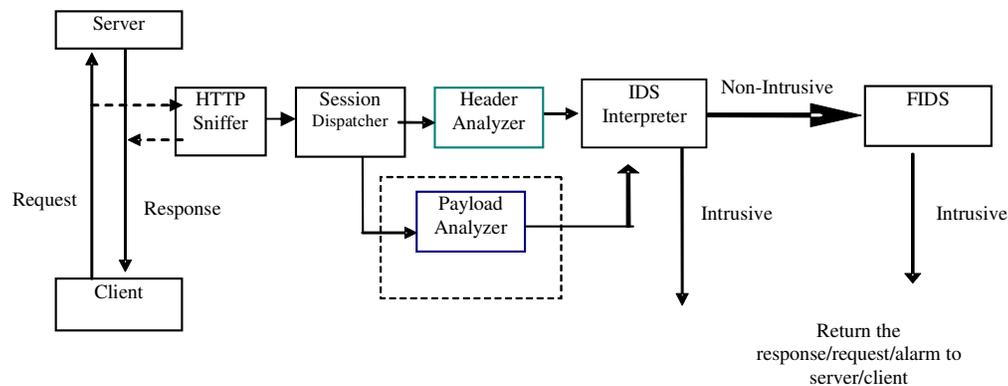

**Figure 1 Block diagram view of integrated FASIDS**

The Header Analyzer reads the header and prepares a list of the objects in the HTTP packets. Each object represents a field of the HTTP protocol and is a five tuple <*message-line, section, feature, operator, content*>. This sequence of objects is given to the IDS interpreter that refers to the rule-base and correlates the different objects to trigger one or more rules. BNF grammar for HTTP in the application layer is designed to cater to the needs of semantic intrusion detection. Simultaneously the Payload Analyzer parses the HTML data and searches for misappropriate usage of tags and attributes and also checks for JavaScript based attacks injected in the HTTP (Mogul et al 1999). The payload analyzer also gives its inputs to the IDS interpreter. The state transition analysis is done by defining states for every match. The incoming pattern is semantically looked-up only in specified states, and this increases the efficiency of the IDS pattern-matching algorithm. If the pattern matches with some predefined pattern then it generates intrusion alert to client/server.

In a Rule-based intrusion detection system, an attack can either be detected if a rule is found in the rule base or goes undetected if not found. If this is combined with FIDS, the intrusions went undetected by RIDS can further be detected. These non-intrusive patterns are checked by the fuzzy IDS for a possible attack. The non-intrusive patterns are normalized and converted as linguistic variable in fuzzy sets.

These values are given to Fuzzy Cognitive Mapping (FCM) (Susan M. Bridges 2002).. If there is any suspicious event, then it generates an alarm to the client/server. FASIDS results show better performance in terms of the detection rate and the time taken to detect. The detection rate is increased with reduction in false positive rate for a specific attack.

## 3.  HTTP RULE-BASE
### 3.1  Header Analyzer

Header Analyzer involves many modules as explained below and depicted in Figure 2.

**HTTP Parser**

After receiving the message from the sniffer, the HTTP parser opens the file and processes it as per the sequence of steps enumerated.

a.  HTTP lexer matches the text from the sniffer with the predefined RFC2616 HTTP specification and provides their equivalent tokens (Mogul et al 1999).

b.  The parser takes the output of the lexer (i.e., the tokens) and checks for violations of specifications and reports if found.

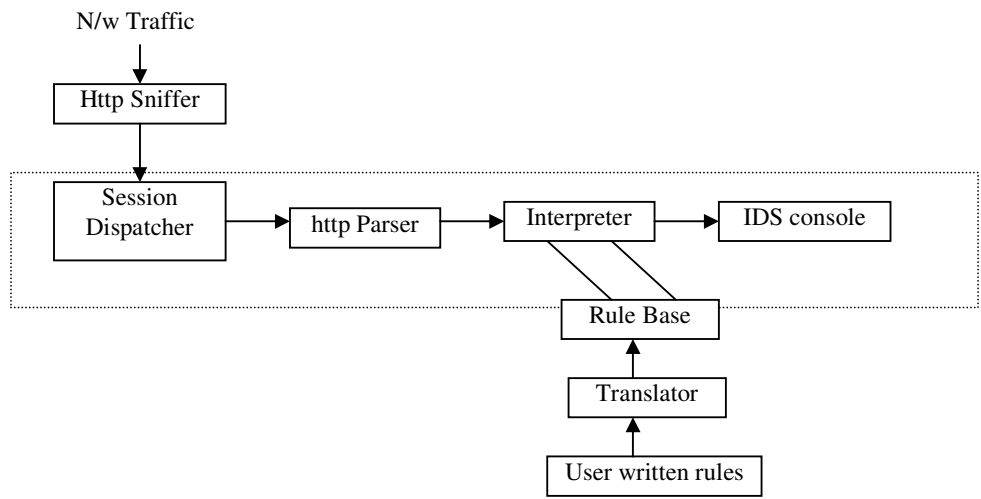

**Figure 2 Header Analyzer**

The output of the HTTP parser is in the form of text and contains various headers and values with a special character ($) as a delimiter. The content of the file is as given below (Line numbers are given for clarity).

```
1…     Request_Method$GET
2…     Request_Request-URI$/index.htm
3…     Request_HTTP-version$HTTP/1.1
4…     generic-header_Accept$image/gif, image/x-xbitmap,
        image/jpeg, image/pjpeg, application/x-
        shockwave-flash, application/vnd.ms-excel,
        application/vnd.ms-powerpoint,
        application/msword, */*
5…     generic-header_Accept-Language$en-us
6…     generic-header_Accept-Encoding$gzip, deflate
7…     generic-header_User-Agent$Mozilla/4.0
        (compatible; MSIE 7.0; Windows NT 5.1; SV1)
8…     generic-header_Host$192.168.0.51:4556
9…     generic-header_Connection$Keep-Alive
```

**Interpreter**

A lexer matches the formatted output from the HTTP parser and state transition analysis is done by defining states for every match. For example, REQLINE_URI is the state that is acquired when the lexer matches with the attribute string 'URI' before its value is encountered. Interpreter refers rule-base and correlates different objects to trigger one or more rules.

**Rule-base**

In the rule-base, each object is associated with a rule as explained below.

**a.    Object**

Each occurrence or a match of a simple elementary pattern is associated with an object which is defined as shown in Table 1.

## Table 1 Object Definition

| Field | May take values |
|---|---|
| Object number | Integer values |
| Message-line | request-line or general-header or body |
| Section | method or uri or version or host or user-agent or content-length |
| Feature | type or size or regex or occurrence |

**b.     Rules**

Each rule contains a set of objects represented by a structure which has *Rule_number, Object_list, No_of_Objects, Message.* A rule will be triggered when a particular set or a sequence of objects occurs.

**Example rules**

Rule 1: Object List = {1, 3, 4};      No. of objects = 3;    In_order= True;
Rule 2: Object List = {2,1,1,1,1};    No. of objects = 5;    In_order= False;
Rule 3: Object List = {3, 6, 4};      No. of objects = 3;    In_order= True;
Rule 4: Object List = {6};            No. of objects = 1;    In_order= False;

**3.1.1   Semantic representation of rules**

Every time a new vulnerability is identified a new rule has to be written for it and is to be updated in the rule-base. The representation of the rules follows the BNF standard with the non-terminals and the terminals being represented in the rule description as shown below in Figure 3.

Rules         :   1*(variable)
Variable      :   message-line section   feature   operator   value
Message-line  :   start-line | header | body
Start-line    :   Request-line | Status-line

| | | |
|---|---|---|
| Header | : | Generic-hdr | Request-hdr | Response-hdr | Entity-hdr |
| Body | : | HTML | XML | … |
| Section | : | Method | Uri | Version | Status-code | <all the generic header fields> |<all the request header fields> |<all the response header fields> |<all the entity header fields> |<all the tags in the HTML doc or other HTTP payload> |
| Feature | : | parameter | size | regex | occurrence |
| Operator | : | = | > | < |
| Value | : | 1*(Alpha | Digit) |
| Alpha | : | [a-zA-Z] |
| Digit | : | [1-9] |

**Figure 3 BNF grammar for HTTP IDS's Rule language**

### 3.2 Payload Analyzer

One of the major limitations in the IDSs that are commercially available is that they do not analyze the payload. It is difficult to do a strict pattern matching over the entire packet in order to figure out the existence of certain information. And if the number of pattern increases, the complexity of the analysis increases. This leads to the necessity to perform protocol analysis (Abbes et al 2004) and apply specific pattern matches as is appropriate for the current part of the packet.

The HTTP Sniffer captures the payload in each thread and stores it in a buffer for the Payload Analysis block as shown in Figure 4. The Payload in individual threads can be HTML documents or Scripts or Images in the web pages.

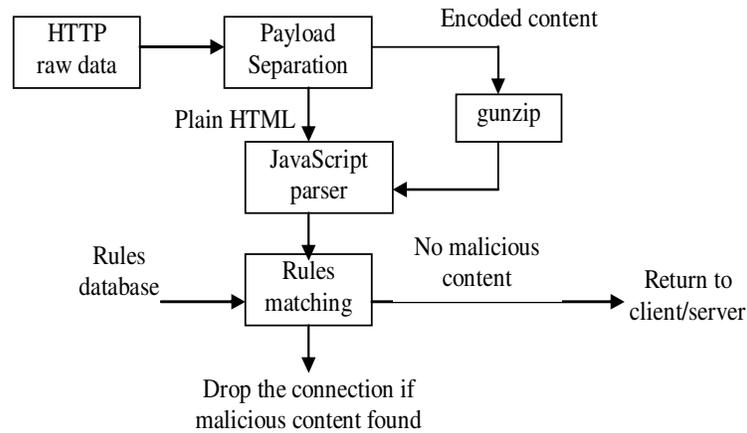

**Figure 4  Functional Block diagram of Payload analysis**

If the payload is zipped, it unzips it and provides the data to the payload analyzer. It parses and analyzes only the HTML and the scripts in the payload and leaves all the Image data, if found. The script content is checked for maliciousness. Table 2 shows the various combinations of Tags and Attributes where Java scripting attacks are generally found. Any suspicious code is reported immediately as an alert.

**Table 2 Specific Tags and Attributes where Java scripting attacks occur**

| TAGS | ATTRIBUTES | CONTENT |
|---|---|---|
| img | Src | Javascript: <some script> |
| anchor | HREF | Javascript: <some script> |
| input | Type | Javascript: <some script> |
| meta | Name | Javascript: <some script> |
|  | Content | Javascript: <some script> |
| div | Align | Javascript: <some script> |
|  | Class | Javascript: <some script> |
| body | Bgcolor | Javascript: <some script> |
|  | Background | Javascript: <some script> |
|  | Leftmargin | Javascript: <some script> |
| iframe | Align | Javascript: <some script> |
|  | Src | Javascript: <some script> |

## 4. FUZZY INTRUSION DETECTION SYSTEM FOR NON-INTRUSIVE PATTERNS

The functional block diagram of fuzzy intrusion detection system is shown in Figure 5. Non-Intrusive patterns are given to the text processor which uses 'awk'. 'awk' is text processing language, which searches for a particular pattern in file and gives the number of occurrences of a particular pattern. The output of this is normalized between 0.0 to 1.0 and goes to fuzzification. Fuzzification converts normalized value into linguistic terms of fuzzy sets. The output of the fuzzification is given to Fuzzy Cognitive mapping which makes use of Fuzzy Associative matrix. This will evaluate all the rules that trigger and uses a mean of maxima defuzzification method to generate its actual response.

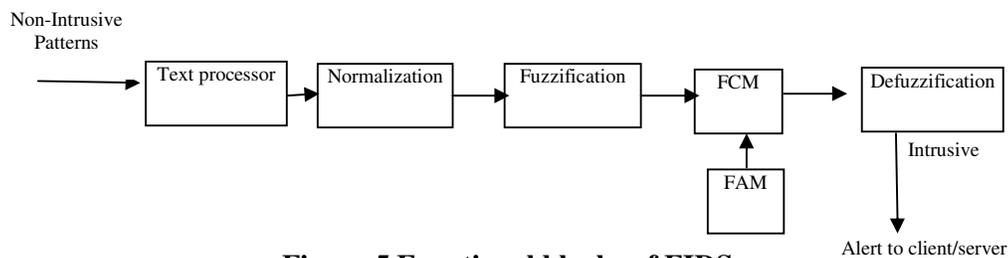

**Figure 5 Functional blocks of FIDS**

### 4.1 Working of Fuzzy Cognitive Mapper in IDS

Several FCM concepts are used for each kind of misuse detection where every "suspicious event" is a kind of misuse detection that affects the machine and user. The events are activated with the help of a fuzzy rule-base where fuzzy rules are used to map multiple inputs. The edges or links are fuzzy relations represented by one or more fuzzy rules. FCM makes use of Fuzzy Associate matrix to evaluate the rules. The output of the FCM again generates an alert to the server/client, if there are any attacks.

### 4.2 Fuzzy Associative Mapping by Fuzzy Rules

Fuzzy Associate Mapping (FAM) is used to map fuzzy rules in matrix format. These rules take two variables as input and map it into two dimensional matrixes. An implementation of this module might use either the matrix or the explicit IF/THEN form. The matrix makes it easy to visualize the system. Fuzzy Associative Mapping shown in Table 3 allows us to conclude the rate of false positives for few attacks such as DoS attacks, Brute force attacks.

**Table 3 Fuzzy Associative Mapping for Brute Force and DoS Attack**

| t \ x | Very Small | Small | Medium | High | Very high |
|---|---|---|---|---|---|
| **Very low** | LP | LP | Non-Intrusive | Non-Intrusive | Non-Intrusive |
| **Low** | LP | LP | LP | Non-Intrusive | Non-Intrusive |
| **Medium** | HP | LP | LP | LP | Non-Intrusive |
| **High** | HP | HP | HP | LP | LP |
| **Very high** | Intrusive | Intrusive | HP | HP | HP |

From the Figure 6, very low, low, medium, high, very high are functions mapping a time scale.

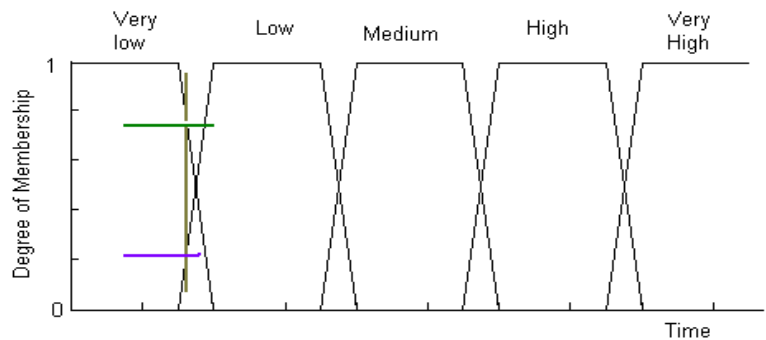

**Figure 6 Fuzzy quantization of trapezoidal-shaped fuzzy numbers**

For particular time, the number of login failure is 0.2 (violet color line) and 0.8 (green color line). It is probably fuzzy at many points. At this point, both rules will trigger. It is low or very low depends on the fuzzy rules, which make use of FAM. If it is more "very low" than it is "low", then "very low" rule will generate a stronger response. The program will evaluate all the rules that trigger and use an appropriate defuzzification method to generate its actual response.

For example, the intruder tries to login with several users' passwords and fails. This attack can be identified by observing the number of login failures and the time interval. FCM for Login_failure is shown in Figure 7.

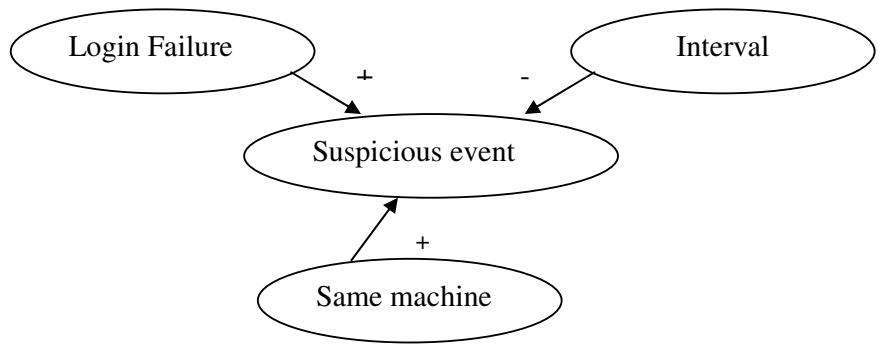

**Figure 7 FCM for Login_failure**

### 4.3   Algorithm for Fuzzy Intrusion Detection System

Step 1 :    Let x= number of login failures and t= time interval

Step 2 :    x=normalization of (x)=(x-*min*)/(*max-min*)
            where,
            *min* is the minimum value for the attribute that x belongs to
            *max* is the maximum value for the attribute that x belongs to

For each numerical attribute, assign the fuzzy space as shown in Figure 6

Step 3 :    The output of the fuzzification is given to FCM which has the fuzzy rules in the form: IF *condition* THEN *consequent*

49

Where,

*Condition* is a complex fuzzy expression i.e., that uses fuzzy logic operators, which is shown in Table 5.2

*Consequent* is an atomic Expression

### Table 4 Fuzzy logic operators

| Logical Operator | Fuzzy Operator |
|---|---|
| x AND t | Min{x,t} |
| x OR t | Max{x,t} |
| NOT x | 1.0-x |

R: IF *y is high* and *t is low*

THEN *Intrusion is HP*

Step 4 : Mean of Maxima defuzzification ($D_{MM}$)=sum $\Sigma x_i$/ |X|

Where, $x_i$ belongs to X

This program will evaluate all the rules that trigger and use a mean of maxima defuzzification method to generate its actual response.

## 5. ATTACKS

There are numerous attacks that can be caused using JavaScript. Cross Site Scripting (CSS), SQL Injection, Denial of Service and Brute force are the most common of all (William Bellamy 2002). The attacks such as Denial of Service and Brute force are not detected by a regular misuse detection system. The fuzzy module adds value by detecting these kinds of attacks.

### 5.1 Cross site scripting attack

A web site may unintentionally include malicious HTML tags or scripts in a dynamically generated page based on invalidated input from untrustworthy sources.

This can be a problem when a web server does not adequately ensure that generated pages are properly encoded to prevent unintended execution of scripts, and when input is not validated to prevent malicious HTML from being passed to the user. By cross-site scripting technique it is possible for an attacker to insert malicious script or HTML into a web page. The purpose of cross-site scripting is that an intruder causes a trusted web server to send a page to a victim's browser that contains malicious script or HTML as desired by the intruder. The malicious script runs with the privileges of a trusted script originating from the trusted web server.

### 5.2     SQL injection attack

Many web pages take parameters from a web user and query the database using SQL. Take for instance when a user logins, the web page asks for user name and password and queries the database to check if a user has valid name and password. With SQL Injection, it is possible for an intruder to send crafted user name and/or password field that will modify the SQL query and thus grant him something else.

### 5.3     Denial of service attack

When a server is intentionally overloaded with many requests from an intruder, it causes it to deny normal access to legitimate users. This attack can also be in the form of an infinite loop that gets executed in the client's browser. The malicious scripts are separated and saved in a text file. It can be given as structured input to the yacc code for signature comparison.

### 5.4     Brute force attack

This attack tries all (or a large fraction of all) possible values till the right value are found, also called an exhaustive search. A brute force attack is a method of defeating a cryptographic scheme by trying a large number of possibilities. For example, exhaustively working through all possible keys in order to decrypt a message. In most schemes, the theoretical possibility of a brute force attack is recognized, but it is set up in such a way that it would be computationally infeasible to carry out.

The output of the rule-based intrusion detection module is non-intrusive for few attacks such as DoS, login failures. In DoS attack, instead of having infinite loop, the intruder will execute the loop for larger number of times. There is a bigger class of attacks which doesn't have a clear rule entry in the rule base can also be detected. These patterns are checked by the fuzzy IDS for a possible attack. Fuzzy Cognitive Mapping is used to capture different types of intrusive behavior as suspicious events.

## 6.    RESULTS

A graph is plotted for the average time taken for scanning a single http request (Response time) versus the number of objects that were incorporated in the IDS interpreter. As the number of objects increase, the number of ways in which the text can be matched increases and hence the time taken also increases. From Figure 8 it can be inferred that the response time increases linearly and then begins to saturate as the number of objects to be matched increases. When the number of objects increases beyond 80, the response time increase at a very slow rate. Hence the implemented IDS perform well when the number of objects is more than 80.

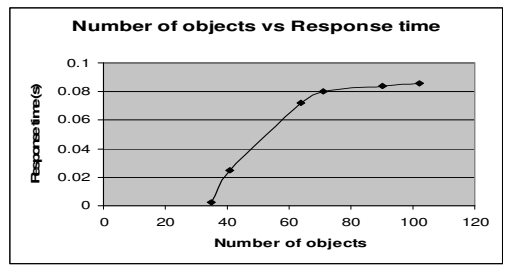

**Figure 8 Performance analysis chart**

The objects in each of the protocol field that are to be searched are plotted in Figure 9. It is observed that if the number of objects to be matched in each protocol field is increasing the Response time increases linearly. But the response time tends to saturate after a specific number of rules. This is because it is expected that the rules contain some common objects which are to be checked once thus improving the response time.

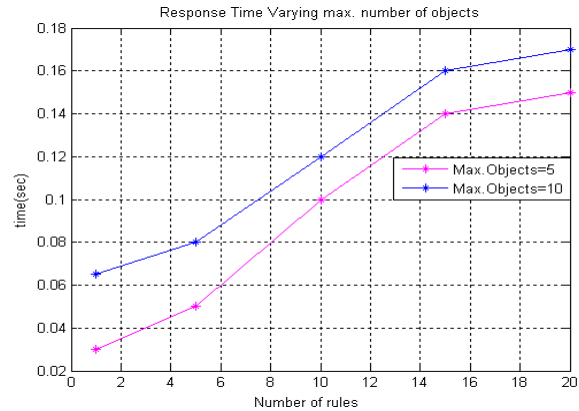

**Figure 9 Response time vs. Rules for each protocol field**

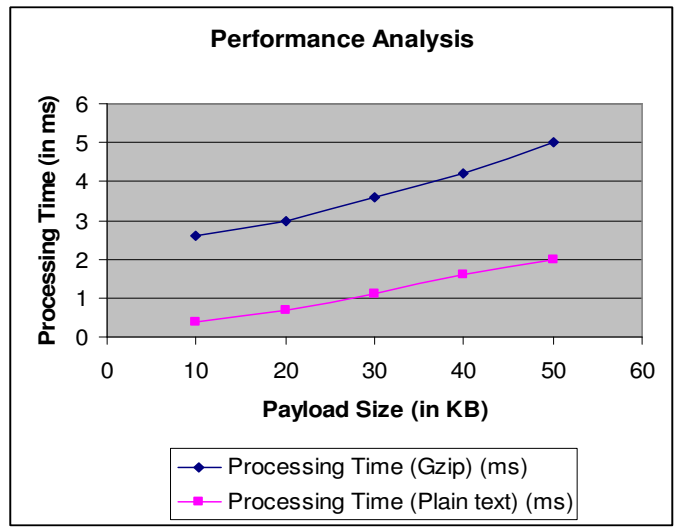

**Figure 10 Performance analyses for payload**

A graph is plotted for the average time taken for analyzing a html file, (Processing time) versus the payload size as shown in Figure 10. As the payload size increases, the amount of the text that needs to be matched increases, and so the processing time also increases.

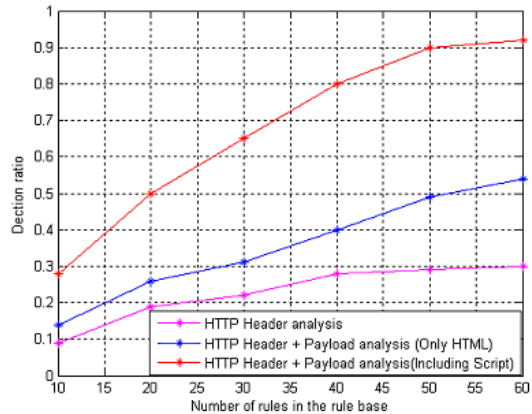

**Figure 11 Detection Ratio with various components of IDS**

Figure 11 shows the detection rate with various components of IDS. From the Figure 6.8, the detection rate increases by combining HTTP header and payload (HTML and Scripts).

Figure 12 shows comparison of Fuzzy based Misuse Detection and Regular Misuse Detection for various attacks. Figure 6 shows the detection rate of fuzzy based misuse detection is high when compared to the regular misuse detection for some attacks such as Dos, brute force, Directory Traversal attacks.

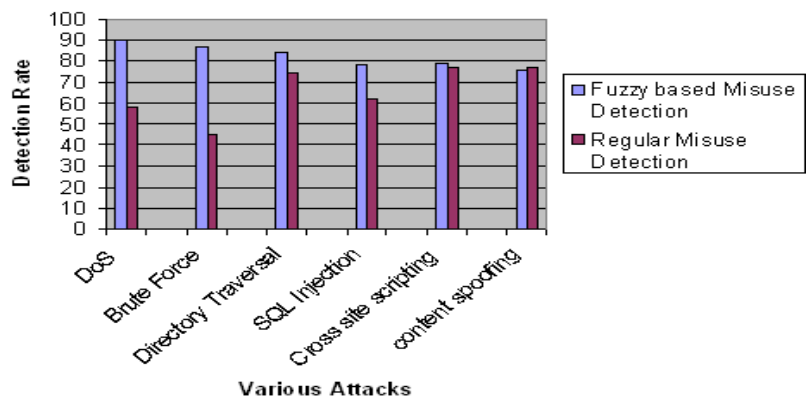

**Figure 12 Comparison of Fuzzy based Misuse Detection and Regular Misuse Detection**

## 7. CONCLUSION

Misuse detection uses rule based IDS that follow a signature-match approach which is time consuming. The rule based semantic intrusion detection system proposed has an efficient memory usage since the amount of memory needed for working of the IDS depends on the rule table size. The IDS developed, will update the signatures and rules automatically, because of the continuously changing nature of attacks. Fuzzy Intrusion Detection System proposed uses Fuzzy Cognitive Mapping (FCM) in order to have an accurate prediction. The complete fuzzy aided application layer semantic intrusion detection system has been implemented in Linux Platform. The system has been tested in web environment and the results are presented. Results show better performance in terms of the detection rate and the time taken to detect an intrusion. The fuzzy aided application layer semantic intrusion detection system can be extended with more semantic parameters by considering protocols other than HTTP carrying payload other than the HTML and JavaScript.

**LIST OF PUBLICATIONS**

1.   Vaidehi V., Srinivasan N., Anand P., Balaji A.P., Prashanth V. and **Sangeetha S.** (2007), 'A Semantics Based Application Level Intrusion Detection System', Proceedings of International Conference on Signal Processing, Networking and Communications, ICSCN 2007, pp. 338-343.

2.   **Sangeetha S.,** Vaidehi V. and Srinivasan N. (2008), 'Implementation of Application Layer Intrusion Detection System using Protocol Analysis', Proceedings of International Conference on Signal Processing, Networking and Communications, ICSCN 2008, pp. 279-284.